\documentclass[12pt]{article}
\usepackage[utf8]{inputenc}
\usepackage[shortlabels]{enumitem}
\usepackage[margin=1.0in]{geometry}
\usepackage{setspace}
\usepackage{lipsum}
\usepackage{longtable}
\usepackage{booktabs} 
\usepackage{makecell}  
\usepackage{float}  
\usepackage[toc,page]{appendix}
\usepackage{rotating}
\usepackage{hyperref}
\usepackage{amsmath}
\usepackage{mathrsfs}
\usepackage{natbib}
\usepackage{graphicx}
\usepackage{mathtools}
\usepackage{scalerel,stackengine}
\stackMath
\usepackage{natbib}
\usepackage[T1]{fontenc}
\usepackage{titling}
\usepackage{caption}
\usepackage{pdfpages}
\usepackage{array}
\usepackage[table]{colortbl}
\usepackage{caption}
\usepackage[svgnames,x11names,table]{xcolor}
\usepackage{pdflscape}
\usepackage{multirow}
\usepackage{subfigure}
\usepackage{caption,tabularx,booktabs}
\usepackage{array}
\usepackage{bbm}
\usepackage{textcomp}
\usepackage[normalem]{ulem}
\useunder{\uline}{\ul}{}

\hypersetup{
    colorlinks=true,
    linkcolor=Navy,    
    urlcolor=blue,     
    citecolor=Navy,     
    filecolor=blue  
}

\begin{document}

\begin{titlepage}
\title{Once Welcomed, Then Scapegoated: \\ The Enduring Consequences of Assimilation Policies in the Wake of Mass Migration}

\author{Vinicius Schuabb\thanks{PhD in Public Policy and Administration, June 2025, Bocconi University, Italy (\href{mailto:vinicius.dinizschuabb@phd.unibocconi.it} {vinicius.dinizschuabb@phd.unibocconi.it}); Research Fellow at the Institute for Mobility and Social Development (IMDS) and Research Associate/Postdoctoral researcher at the Center for Empirical Studies in Economics (FGV CEEE), Rio de Janeiro, Brazil.} \\ 
    \\
    \textit{Job Market Paper}\thanks{I am grateful to Maristella Botticini (Bocconi University) for her supervision and guidance. I thank Laura Ogliari and Giampaolo Lecce (University of Bergamo) for helpful comments, and Vincenzo Galasso, Paolo Pinotti, Manoel Gehrke, Henrique Alpalhão, Paolo Agnolin, Elena Marseglia, and Gian Maria Mallarino (Bocconi University) for valuable suggestions. I also thank Claudio Ferraz (UBC and PUC-Rio), Pedro Américo (USP), and Leonardo Monasterio (IPEA) for insightful feedback on earlier stages of this project. Finally, I thank the Graeff family for the inspiration provided during my exchange semester in Karlsruhe, Germany, in 2012.}
    }

\date{For the most up-to-date version, see: \href{https://arxiv.org/abs/2509.02836}{arxiv.org/abs/2509.02836}}

\maketitle


\begin{abstract}
\noindent This paper examines the short- and long-term effects of immigrant assimilation policies in Brazil after the Mass Migration period. It focuses on the Nationalization Campaign, launched amid rising anti-immigrant sentiment during the Great Depression and the World Wars. Using newly assembled archival data spanning the twentieth century, I assess the campaign’s impact on education among immigrants and their descendants following the closure of immigrant-community schools. In the short term, the policy sharply reduced educational attainment among targeted groups; in the long term, school-age immigrants during the campaign completed less education over their lifetimes, with adverse effects persisting into the second generation. The magnitude of these impacts varied with cultural proximity to native Brazilians.

\end{abstract}



\setcounter{page}{0}
\thispagestyle{empty}
\end{titlepage}
\pagebreak
 \newpage

\doublespacing

\newpage

\section{Introduction} \label{sec:Introduction}

\noindent Immigration remains a disputable topic in the twenty-first century, fueling ongoing debates over integration and assimilation worldwide \citep{oecd_2017}. In response, scholars have renewed their interest in the Age of Mass Migration (1850–1930). Understanding how millions of Europeans and Asians resettled across the Americas offers valuable insights into the persistent challenges faced by migrants and host societies, insights that remain highly relevant for contemporary policy debates.

This research investigates the long-term consequences of assimilation policies implemented during Brazil's Nationalization Campaign (NC), following the Age of Mass Migration, with a particular focus on the experiences of immigrant communities. It examines how changes in the educational system affected the human capital development of school-age immigrants and their descendants. To identify these effects, I leverage the timing of major historical events, as the Great Depression and the two World Wars, alongside shifts in Brazilian government policies, including changes in immigration regulation. I employ a variety of empirical strategies at the state, municipal, and individual levels to assess differential exposure to these policies. Comparison groups are defined by the historical presence of immigrants from specific countries of origin and the timing of their arrival in Brazil.

In the short term, the Nationalization Campaign had significant negative effects on the educational system, including declines in private schools, used as a proxy for immigrant-community institutions, as well as in student enrollment and teacher availability at the state level. Municipalities with a higher concentration of targeted immigrant populations and rural schools, where immigrant communities were typically located, did also experience lower literacy rates and reduced primary and secondary school completion between the 1940 and 1950 Censuses. Immigrant workers also faced higher unemployment rates, while the participation of immigrant teachers in local labor markets increased.

In the long term, immigrants of school age during the NC are compared to similar cohorts who arrived in Brazil after the campaign, with no significant differences between groups at the time of arrival. Individual-level data from the 1991, 2000, and 2010 Censuses reveal a persistent negative effect of NC exposure on educational attainment for both targeted immigrants and their descendants, indicating an intergenerational transmission of the assimilation policies.

A key mechanism driving these effects was cultural proximity, as highlighted by \citet{lecce_et_al_JEG_2022}, \citet{fouka_et_al_RES_2021}, and \citet{tabellini_RES_2019}. The Nationalization Campaign promoted a narrow vision of nationalism centered on the Portuguese language and Catholicism. Empirical evidence supports historical accounts indicating that immigrants from culturally distant backgrounds faced greater barriers to adaptation and, in some cases, actively resisted these norms. Japanese and German communities, for example, were more severely affected than Italian ones, particularly in cultural and educational domains. Cultural change tends to be gradual and often meets resistance, especially when majority and minority cultures diverge significantly \citep{giuliano_nunn_2021, carvalho_koyama_2016}. In this context, cultural and genetic proximity may condition the effectiveness of assimilation policies and help explain variation in immigrant integration outcomes \citep{arbatli_et_al_ECON_2020, spolaore_wacziarg_QJE_2009}.

The foundations of this process lie in concerns with demographic composition and cultural cohesion, ideas central to political thought since Plato and Aristotle, who viewed education as essential to the formation of the \textit{polis}.\footnote{See, for example, Plato's Republic, Laws, and Statesman.} Modern nation-building processes have similarly relied on assimilation policies, especially educational reforms, as tools to foster institutional trust and national identity \citep{alesina_2021a, alesina_la_ferrara_2002, lott_1999, hall_1992, hobsbawm_1992}. Authoritarian regimes, in particular, have strong incentives to pursue such policies,\footnote{``... the diminishing returns to controlling the citizenry through force increases the marginal return to indoctrination.'' \citep{lott_1999}} often adopting measures aimed at fostering trust between minority groups, the broader population, and national institutions \citep{alesina_la_ferrara_2002}.

Brazil's Nationalization Campaign exemplifies how state-led assimilation policies can shape long-term outcomes. This context provides a valuable setting to examine the intergenerational effects of forced assimilation on human capital formation across diverse immigrant groups \citep{cunha_heckman_AER_2007, becker_tomes_1986}. Since intergenerational cultural transmission plays a key role in the accumulation of skills and knowledge \citep{bisin_rubin, bisin_verdier}, assimilation policies may have led to persistent gaps in educational attainment \citep{squicciarini_2020, botticini_eckstein, becker_woessmann_2009}. These disparities may, in turn, affect long-run local development, as immigrant-linked human capital has been shown to yield lasting economic benefits \citep{rocha, becker_2011}.

This analysis contributes to our understanding of the mechanisms driving intergenerational inequality and local development under state-led immigration and education policies \citep{abramitzky_2021, bloome_2018, torche_2010}. It also adds to the growing literature on assimilation efforts implemented through educational reforms, particularly those involving the imposition of a national language of instruction \citep{fouka_2020, lleras_AEJEC_2015, eriksson_EHDR_2014, clots_et_al_EJ_2013}.

The remainder of the paper is organized as follows. Section \ref{sec:Historical Context} presents the Brazilian historical context. Section \ref{sec:Data}, the data used for the analysis. Section \ref{sec:Short-term effects} the short-term analysis and results. Section \ref{sec:Intergenerational effects}, the long-term, intergenerational analysis and results. Finally, Section \ref{sec:Discussion and Conclusion} a discussion about the findings and conclusions.

\section{Historical Context} \label{sec:Historical Context}

\noindent The Mass Migration of the nineteenth and twentieth centuries brought a diverse array of ethnic groups to the Americas \citep{sanchez_alonso_2018, acosta_2018, fitzgerald_2014}, with Brazil receiving nearly 5 million European and Asian immigrants between 1819 and 1947, one of the largest migration episodes in modern history \citep{balderas_greenwood_jpe_2010}. Many of these immigrants settled in the economically developing southern regions of Brazil, particularly in remote rural areas of states like Rio Grande do Sul and São Paulo, where they established largely autonomous communities. Within these communities, immigrant-run schools were central institutions, serving to transmit language, religion, and cultural traditions across generations, especially in areas underserved by the national education system \citep{kreutz_2010}. For instance, by 1940, Brazil counted 644,255 German-speaking and 458,093 Italian-speaking residents.

However, global and domestic upheavals, including the Great Depression and two World Wars, reshaped public opinion, and immigrants were increasingly scapegoated for economic and political instability. In response, the Brazilian government introduced assimilationist policies, including educational reforms and immigration quotas. Although local enforcement varied, by the 1930s there were still approximately 2,500 immigrant community schools, about 7\% of all schools nationwide - including 1,579 German, 396 Italian, 349 Polish, and 178 Japanese institutions \citep{kreutz_2000}.

\subsection{Migration Policies}

The Great Depression of 1929 triggered a severe economic crisis in Brazil, where coffee exports—central to the national economy, lost 70\% of their value. In the wake of this collapse, political leaders such as Getúlio Vargas began to scapegoat immigrants, accusing them of taking jobs from native Brazilians. Immigrants, once welcomed to replace enslaved labor, were now portrayed as contributing to national instability. Rising xenophobia during the 1930s and 1940s coincided with growing public support for restrictive measures, reshaping the state’s approach to immigration.

This shift translated into a series of legal and institutional changes. The 1930 Presidential Decree No. 19,482—the ``Two-Thirds Act''—barred ``third-class'' foreigners and required firms to hire at least two-thirds Brazilian nationals. In 1934, nationality-based quotas further restricted immigration, allowing only 2\% of each nationality’s immigration total over the previous 50 years—favoring groups like Italians and Portuguese. The process culminated in the 1938 Nationalization Campaign, including Decree No. 3,010, which sought to regulate the entry and stay of foreigners in Brazil, their distribution and assimilation. The decree emphasized preserving Brazil's ethnic composition, political structures, and economic and cultural interests. While severe, these policies remained largely in place until the 1964 military coup. I use this continuity to compare cohorts exposed to the Nationalization Campaign with later arrivals, using 1963 as a cutoff to assess the long-term impact of these assimilationist policies.

\subsection{Reforms in the educational system}

As part of his effort to unify the Brazilian population around a national ideal of citizenship, Vargas used the educational system as a central instrument of the Nationalization Campaign. The campaign accelerated earlier educational reforms, beginning with the 1931 Francisco Campos Reform and culminating in the 1942 Capanema Reform, which restructured curricula to prioritize patriotism and vocational training while creating a new institutional framework, including the National Board of Education. A core target of these reforms was the immigrant-community schools: beginning with Decree No. 406 in 1938 and culminating in Decree No. 1545 in 1939, the state ordered the closure of hundreds of such schools, imposed Portuguese as the sole language of instruction, and restricted immigrants' access to teaching jobs. These measures later extended to banning foreign-language newspapers, radio broadcasts, books, and even religious and cultural ceremonies. 

Historical accounts suggest that these policies were often brutally enforced, with military and state agents using intimidation, arrests, and violence to ensure compliance \citep{kreutz_2010}. Wartime rhetoric further fueled this agenda, especially after Brazil joined the Allies in 1942; immigrant communities—particularly German, Italian, and Japanese—were portrayed as potential enemies. A general's infamous remark, ``It is better to raise ignorant people than traitors,''\footnote{General Meira de Vasconcellos: ``Antes criarmos ignorantes que criarmos traidores'' cited in Bethlem. H. 1939. \textit{O Vale do Itajaí. Jornadas de Civismo}.} captured the regime’s position. As immigrant schools, often secular and community-oriented, were dismantled, they were slowly replaced by under-resourced Brazilian public schools, suggesting that assimilation policies likely imposed both quantitative and qualitative educational setbacks on immigrant children \citep{squicciarini_2020}.

\begin{figure}[H]
    \centering
    \caption{Number of students in primary education - Brazil}
    \includegraphics[width=13cm]{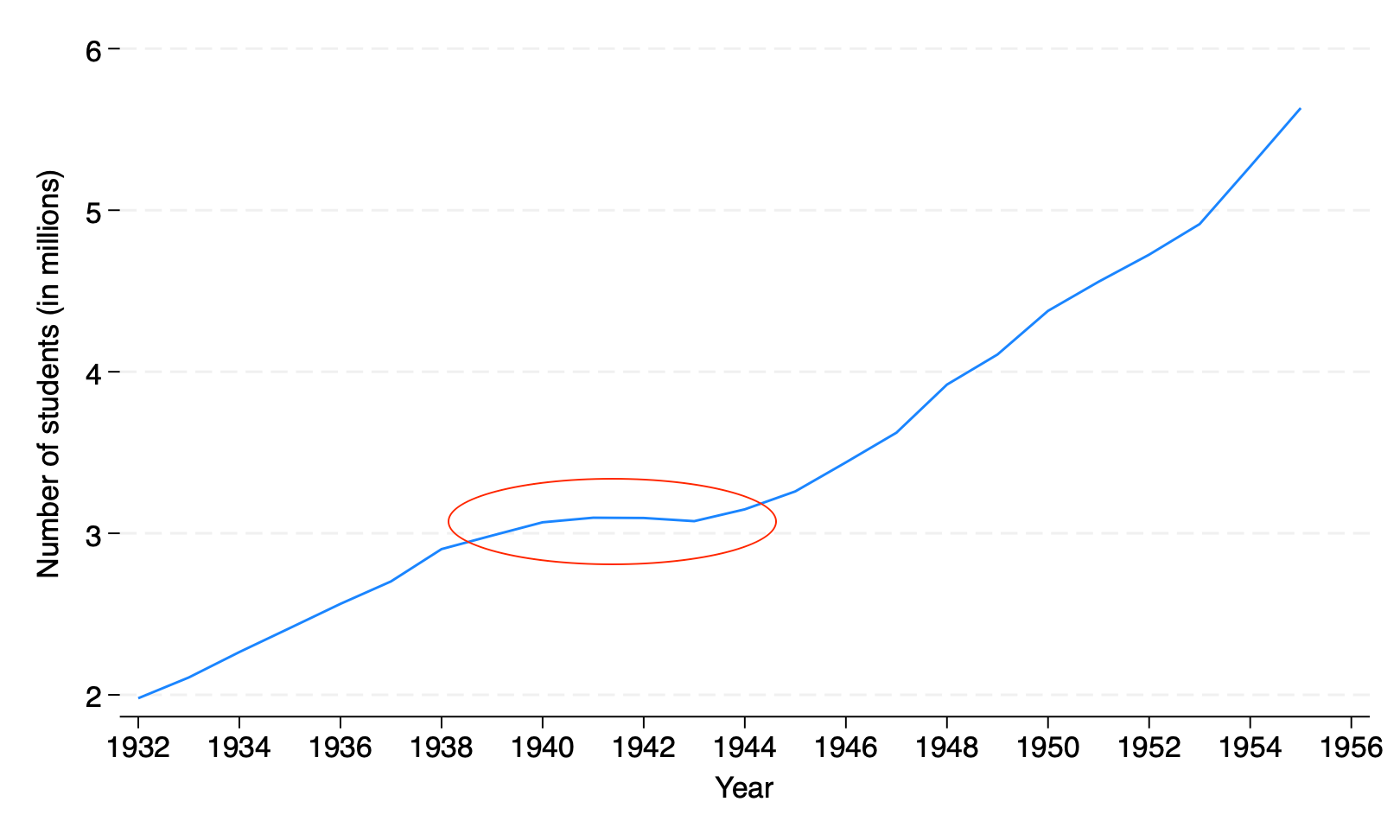}
    \label{fig:n_students_primary_br}
    \caption*{\footnotesize Source: Data from the very first annual reports on the educational system published by the Brazilian Ministry of Education (INEP/MEC) and compiled by the Brazilian Institute of Geography and Statistics (IBGE).}
\end{figure}

Empirical evidence corroborates historical accounts. Figure \ref{fig:n_students_primary_br} depicts the number of primary school students in Brazil from 1932 to 1955, revealing a pronounced structural break during the Nationalization Campaign in the late 1930s and early 1940s. This period marks a clear interruption in the previously steady growth of student enrollment, which abruptly stalled and remained stagnant. While \citet{kang_2017} identified this disruption in his analysis of educational outcomes between 1930 and 1964, the underlying causes remained ambiguous. I contend that the assimilation policies enacted as part of the Nationalization Campaign played a central role in driving this educational stagnation.

\subsection{Diversity Among Immigrant Communities}

Census data reveal the diverse origins of the ethnic groups that migrated to Brazil, with some of the largest communities being of Italian, Japanese, and German descent. These groups became prime targets of the Nationalization Campaign due to their perceived ties to the Axis powers during World War II. However, historical evidence shows that the campaign's reach extended beyond them — immigrant communities of Slavic origin, among others, also faced significant repercussions from the widespread enforcement of assimilation policies.

Historical and anthropological studies indicate that the severity of assimilation policies' impact on education and religious practices varied according to each immigrant group’s cultural proximity to the state’s prescribed ideal of “Brazilianness.”  Immigrants from Southern European countries — especially Italy, Portugal, and Spain — were generally less affected, while those from more culturally distant backgrounds, such as Germany, Japan, and Poland, faced stricter enforcement and deeper cultural disruption. Notably, although Italian immigrants were also targeted by these policies, their cultural and linguistic similarities with the dominant national identity facilitated a comparatively smoother assimilation process.

Supporting this, Table \ref{tab:language_home_1940} shows that only 16\% of Italian and 21\% of Spanish immigrant households reported using their mother tongue at home, compared to 58\% of German and 85\% of Japanese households. The enforcement and impact of assimilation policies were mediated by such cultural proximity factors. Communities that more readily embraced ``Brazilianness''—through adopting Portuguese language use, practicing Catholicism, and conforming to other cultural norms—suffered less disruption than those whose cultural backgrounds diverged more sharply from the national ideal.

\begin{table}[H]
\centering
\caption{Immigrant households use of mother tongue at home - 1940 Brazilian Census}
\resizebox{0.6\textwidth}{!}{%
\begin{tabular}{lc}
\toprule
\textbf{Country of Origin} & \textbf{Preferably spoke mother tongue} \\ \hline
\textbf{Japanese} & \textbf{85\%} \\
\textbf{German} & \textbf{58\%} \\
Russian & 53\% \\
Polish & 48\% \\
Austrian & 42\% \\
Spanish & 21\% \\
\textbf{Italian} & \textbf{16\%} \\
\bottomrule
    \end{tabular}
}
\caption*{\footnotesize Source: Author with data from the Brazilian 1940 Census.}
\label{tab:language_home_1940}
\end{table}

\section{Data} \label{sec:Data}

\noindent This study draws on historical administrative data at the state, municipal, and individual levels. I construct panel datasets by merging multiple twentieth-century sources, detailed as follows.

At the state level, educational data come from early publications of the Ministry of Education and Health (MEC) and the Brazilian Institute of Geography and Statistics (IBGE), including records I digitized from historical archives. The dataset spans 1932 to 1955 and includes information on the number of schools, students, teaching staff (by origin), and student outcomes.\footnote{Data source: Brazilian Ministry of Education and Health, Sinopse Regional do Ensino Primário Fundamental Comum – Dados Retrospectivos 1940–1957. Due to missing or preliminary records, 1956 and 1957 are excluded.} Although specific data on immigrant community schools are unavailable, I exploit their predominantly private administration \citep{kreutz_2000} to examine variation in private school outcomes across regions with high immigrant concentrations.\footnote{Between 1940 and 1947, private schools fell by 2,416, close to the 2,500 immigrant schools in the 1930s \citep{kreutz_2010}, suggesting targeted suppression.}

At the municipal level, demographic and socioeconomic data come from the 1940 and 1950 Censuses, which provide information on age, sex, citizenship, place of origin, and education. As these records are only available as scanned images, I applied Optical Character Recognition (OCR) to extract the data and constructed a municipality-level panel for both years. Given the campaign’s focus on rural, immigrant communities, I use cross-municipality variation to estimate its local effects.

Finally, I use three individual-level sources. The first is a sample of approximately 100 randomly selected passenger lists from ships arriving at the Port of Santos (1930–1963), from which I extract information on immigrants' age, literacy, and occupation at arrival. The second consists of worker registration forms from Rio Grande do Sul (1930–1944), which record requests for work permits and include data on immigrants' age, origin, and occupation. The third is microdata from the 1991, 2000, and 2010 Brazilian Censuses. I use the 10\% household samples provided by the \textit{Centro de Estudos da Metrópole} (CEM-USP) to analyze the long-term outcomes of immigrants who arrived between 1930 and 1963.


\section{Short-term effects of the assimilation policies} \label{sec:Short-term effects}

The analytical section begins by quantifying the short- and medium-term effects of the Nationalization Campaign on educational outcomes. Results are presented in a macro-to-micro sequence, exploiting variation in exposure to assimilation policies across states, municipalities, and individuals. At the state level, I estimate differential trends in educational indicators across regions with varying concentrations of immigrant groups by origin. At the municipal level, I compare changes in educational outcomes between the 1940 and 1950 Censuses for municipalities in Rio Grande do Sul to capture localized treatment effects. At the individual level, I analyze shifts in formal labor market participation in Rio Grande do Sul, focusing on occupational outcomes among immigrant teachers and unemployed workers. Taken together, these analyses yield consistent evidence of how targeted assimilation policies disrupted educational attainment trajectories.

\subsection{State-Level: Educational Trends and Policy Exposure} \label{subsec:short_state_level}

The state-level analysis draws on annual educational data from 1932 to 1955, gathered from historical archives. It leverages cross-state variation to identify changes in educational outcomes associated with the immigrant groups targeted by the Nationalization Campaign.

\subsubsection{Empirical Strategy: Dynamic Difference-in-Differences}

\noindent This analysis employs a difference-in-differences strategy to estimate the dynamic effects of the Nationalization Campaign on educational outcomes. The identification strategy relies on the assumption that, prior to the campaign, states followed comparable trends in educational indicators. States with higher concentrations of non-Southern European immigrants, however, were more directly affected by the campaign, particularly through the targeted closure of immigrant-community schools.

\begin{figure}[H]
    \centering
    \caption{Trends pre- and post-treatment: number of private schools in Brazilian states 1932-1955}
    \includegraphics[width=15cm]{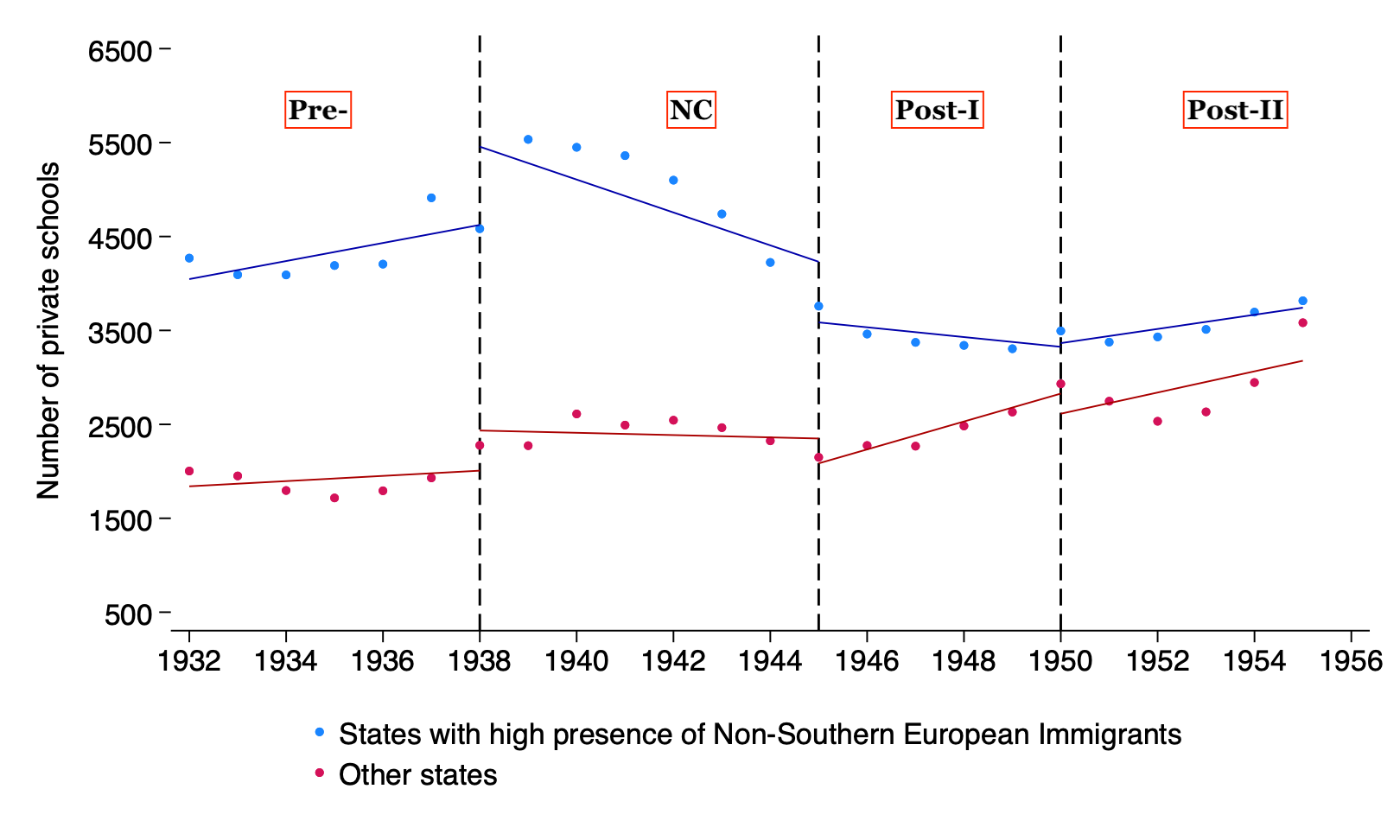}
    \label{fig:trends_did_pre_post}
    \caption*{\footnotesize Source: Data from the very first annual reports on the educational system published by the Brazilian Ministry of Education (INEP/MEC) and compiled by the Brazilian Institute of Geography and Statistics (IBGE). Immigrants from non-Southern European origins are Asians and Europeans not from Italy, Spain, and Portugal; mainly Japan, Germany, and Poland.}
\end{figure}

Figure \ref{fig:trends_did_pre_post} presents pre- and post-treatment trends in the number of private schools, used here as a proxy for immigrant-community schools.\footnote{For statistical purposes, immigrant-community schools were classified as private institutions \citep{kreutz_2000}.} The analysis compares two groups of states: those with a high concentration of non-Southern European immigrants, who were the main targets of the Nationalization Campaign, and all others. Treatment timing follows changes in federal administrations, based on the periodization proposed by \citet{kang_2017}.

Prior to the campaign (Pre-), both groups exhibited broadly similar trends. Between 1938 and 1945 (NC), however, only the more heavily targeted states experienced a marked decline in private, immigrant-community schools. During the first democratic period following Vargas’s rule (Post-I), non-targeted states saw a sharp increase in private school numbers, possibly driven by internal migration. By the second post-treatment period (Post-II), both groups had largely resumed their pre-campaign trajectories.

The short-term effects of the Nationalization Campaign on educational outcomes are estimated using a state-level panel dataset, applying the following reduced-form specification:

    \begin{equation*} 
    \begin{aligned} 
        \text{Education}_{s,t} = \; & \alpha + \beta_1 \;(Non{\text -}Southern\;European\;Immigrants_{s, 1940} \;* \; \mathbbm{1}_{1938-1945}) \; + \\
        &  + \beta_2 \;(N{\text -}SEI_{s, 1940} \;* \; \mathbbm{1}_{1946-1950}) + \beta_3 \;(N{\text -}SEI_{s, 1940} \;* \; \mathbbm{1}_{1951-1955}) \; + \\
        & \; + \; \rho_{s} + \lambda_{t} + \epsilon_{s,t}
    \end{aligned}
    \end{equation*}

\noindent where, \textit{Education} denotes the educational outcomes observed in state \textit{s} and year \textit{t}. The term \textit{Non-Southern European Immigrants} is computed either as the logarithm of the non-Southern European immigrant population in state \textit{s} (intensive margin) or as a binary indicator for states that historically received substantial inflows of immigrants from those origins (extensive margin),\footnote{States with significant non-Southern European immigrant populations include Rio Grande do Sul, Paraná, Santa Catarina, São Paulo, Mato Grosso, Distrito Federal (Capital), Rio de Janeiro, Minas Gerais, and Espírito Santo.} based on the 1940 Census.\footnote{The 1940 Census provides the most proximate demographic snapshot to the onset of the Nationalization Campaign, with most data collected during the pre-treatment period.} The terms $\mathbbm{1}_{1938\text{-}45}$, $\mathbbm{1}_{1946\text{-}50}$, and $\mathbbm{1}_{1951\text{-}55}$ are indicator variables for successive treatment periods,\footnote{See \citet{wolfers_AER_2006} for a similar dynamic difference-in-differences setup.} capturing the timing of the Nationalization Campaign and its aftermath. The parameters $\beta_i$ represent the coefficients of interest, measuring the interaction effects between exposure to the policy and each treatment period. The specification includes state fixed effects ($\rho_s$) to control for time-invariant state-level heterogeneity and year fixed effects ($\lambda_t$) to account for nationwide shocks. The idiosyncratic error term $\epsilon_{s,t}$ is clustered at the state level.

\subsubsection{Results: Educational Outcomes and Targeted Immigrant Groups}

Table \ref{tab:did_short_results_poisson} reports Poisson panel difference-in-differences estimates at the state level, focusing on the interaction between treatment periods and measures of treatment intensity. The first panel uses the log of the immigrant population from the 1940 Census as a continuous treatment variable (intensive margin), while the second uses a binary indicator for historically targeted states (extensive margin). Results reveal a significant and growing negative impact of the Nationalization Campaign on the number of students, schools, and teachers over time. Notably, the effects extend beyond raw enrollment and staffing: as the treatment accumulates, it begins to significantly reduce student pass rates. The decline in school numbers is primarily driven by closures of private institutions, used here as a proxy for immigrant-community schools, while the drop in teachers stems mainly from those trained outside the public system, likely immigrant educators. In terms of magnitude, the intensive margin estimates indicate that a 1\% increase in the number of non-Southern European immigrants is associated with a 13\% decline in private schools by the final period (Post-II). The extensive margin results show that being a state with a high presence of targeted immigrants is associated with a 62\% reduction in private schools over the same period. These results align closely with historical accounts of the Nationalization Campaign's disruptive impact on educational access in immigrant-dense regions.\footnote{Pre-treatment (1932–1937) corresponds to Vargas's first republican term; the Nationalization Campaign (1938–1945) to his dictatorship; 1946–1950 to President Dutra's administration; and 1951–1955 to Vargas’s second presidency, ending in 1954 with his vice president assuming office.}

As a robustness check, \autoref{tab:did_short_results_ols} in \autoref{sec:appendix} reports OLS estimates of the short-term state-level analysis, using logged dependent variables to account for skewed distributions. The results closely mirror those from the Poisson specifications in both magnitude and significance, reinforcing the robustness of the main findings.


\bigskip

\begin{table}[H]
\centering
\caption{Effects of the Nationalization Campaign on educational outcomes given the prevalence of non-Southern European immigrants in Brazilian States}
\resizebox{1.0\textwidth}{!}{%
\begin{tabular}{lcccccc}
\hline
Dependent Variable (N) & Students & Approved Students & Schools & Private Schools & Teachers & Untrained Teachers \\ \hline
 & \multicolumn{1}{l}{} & \multicolumn{1}{l}{} & \multicolumn{1}{l}{} & \multicolumn{1}{l}{} & \multicolumn{1}{l}{} & \multicolumn{1}{l}{} \\
\multicolumn{7}{c}{\textit{Intensive Margin: log(Non-Southern European Immigrants) x treatment period}} \\
$N\text{-}SEI * \mathbbm{1}_{1938-1945}$ (NC) & -0.009 & -0.003 & -0.012 & -0.006 & -0.017** & -0.060*** \\
$N\text{-}SEI * \mathbbm{1}_{1946-1950}$ (Post-I) & -0.030** & -0.028 & -0.039** & -0.103*** & -0.038*** & -0.111*** \\
$N\text{-}SEI * \mathbbm{1}_{1951-1955}$ (Post-2) & -0.035** & -0.064** & -0.065*** & -0.130*** & -0.047*** & -0.160*** \\
 & \multicolumn{1}{l}{} & \multicolumn{1}{l}{} & \multicolumn{1}{l}{} & \multicolumn{1}{l}{} & \multicolumn{1}{l}{} & \multicolumn{1}{l}{} \\
\multicolumn{7}{c}{\textit{Extensive Margin: high Non-Southern European immigrant presence (dummy) x treatment period}} \\
$N\text{-}SEI * \mathbbm{1}_{1938-1945}$ (NC) & -0.082** & -0.082 & -0.068 & -0.128 & -0.087 & -0.232*** \\
$N\text{-}SEI * \mathbbm{1}_{1946-1950}$ (Post-I) & -0.167*** & -0.189 & -0.130 & -0.535** & -0.143 & -0.295** \\
$N\text{-}SEI * \mathbbm{1}_{1951-1955}$ (Post-2) & -0.205*** & -0.379*** & -0.292*** & -0.623*** & -0.224** & -0.485*** \\
 & \multicolumn{1}{l}{} & \multicolumn{1}{l}{} & \multicolumn{1}{l}{} & \multicolumn{1}{l}{} & \multicolumn{1}{l}{} & \multicolumn{1}{l}{} \\
N of observations & 528 & 528 & 528 & 528 & 528 & 528 \\
N of unique states & 22 & 22 & 22 & 22 & 22 & 22 \\
State and Year FEs & X & X & X & X & X & X \\ \hline
\end{tabular}
}
\caption*{\footnotesize Source: Author. Results of a Poisson panel difference-in-differences estimates with dynamic treatment. The treated states are those with high prevalence of non-Southern European immigrants, given by the log number of immigrants minus the number of Portuguese, Spanish, and Italian immigrants in the population of a state in the 1940 Brazilian Census. Treatment periods of time are historically defined as federal government changes. The pre-period represents the first republican rule of Getúlio Vargas; the Nationalization Campaign period, his rule as a dictator; Between 1946 and 1950, the rule of elected president Gaspar Dutra; then, Vargas again as elected president between 1951 and 1955 (although he died in 1954 and his vice took office for 1955). *** p$<$0.01, ** p$<$0.05, * p$<$0.1}
\label{tab:did_short_results_poisson}
\end{table}

\subsection{Municipal-level: Local Variation in Educational Outcomes Over Time}\label{subsec:short_municipal_level}

The state-level results suggest significant and sizable effects of the Nationalization Campaign's assimilation policies on educational outcomes in targeted states. To explore local variation more closely, I now turn to a comparison between Censuses at the municipal level.

\subsubsection{Empirical Strategy: Interaction with Information on Rural Schools}

In the municipal-level analysis, I examine changes in educational outcomes between the 1940 and 1950 Brazilian Censuses for municipalities in the state of Rio Grande do Sul. Given that many targeted, immigrant-community schools were located in rural areas \citep{kreutz_2000}, exposure to the Nationalization Campaign is measured through the interaction between the presence of non-Southern European immigrants and the number of rural schools in 1940. The reduced-form estimation identifies the relationship between changes in the local educational system and exposure to assimilation policies using the following specification:

    \begin{equation*}
    \begin{aligned}
        \Delta \; Educ._{m,1950-1940} = \; & \alpha + \beta \; (Non{\text -}Southern\;European\;Imm._{m, 1940} \;* \; \text{Rural Schools}_{m, 1940}) \; + \\ 
        & + \delta \; Non{\text -}Southern\;European\;Imm._{m, 1940} + \rho \; Rural \; Schools_{m, 1940} \; + \\
        & + X'_{m,1940} \; \Lambda + \epsilon_{m,t}
    \end{aligned}
    \end{equation*}

\noindent where, $\Delta$\textit{Educ.} denotes the change in educational outcomes in municipality \textit{m} between the 1940 and 1950 Censuses. \textit{Non-Southern European Immigrants} is the log number of immigrants from non-Southern European origins in municipality \textit{m} as reported in the 1940 Census, and \textit{Rural Schools} is the number of rural schools in the same year. The coefficient of interest, $\beta$, captures levels of the treatment effect by associating the presence of non-Southern European immigrants and the number of rural schools with changes in local educational outcomes. The vector $X'$ includes control variables from the 1940 Census capturing municipal characteristics, and $\epsilon_{m,t}$ is a robust, idiosyncratic error term.

By specifying the dependent variable in first differences, the model captures within-municipality changes over time, effectively controlling for unobserved, time-invariant characteristics. This approach isolates the association between exposure to the Nationalization Campaign and changes in local educational outcomes, such as school closures or reductions in teaching staff, highlighting how these shifts were influenced by the presence of targeted immigrant populations and the rural distribution of immigrant-comunity schools.

\subsubsection{Results: Educational Changes Between the 1940 and 1950 Censuses}

Table \ref{tab:munic_results} presents OLS estimates using changes in local characteristics between the 1940 and 1950 Censuses as dependent variables. The interaction between the number of rural schools and the log of non-Southern European immigrants is negatively and significantly associated with educational outcomes. Consistent with the declines in approved students, teachers, and schools reported in Table \ref{tab:did_short_results_poisson}, the results indicate sizable reductions in the shares of literate individuals, primary school completers, and high school graduates in municipalities more exposed to the Nationalization Campaign. These patterns may also reflect another dimension of the campaign: policies that encouraged native Brazilians to move to immigrant settlements, which may have simultaneously triggered outmigration of immigrants and their descendants from these areas as well.

\begin{table}[H]
\centering
\caption{Changes in characteristics of municipalities between the 1950 and 1940 Census - State of Rio Grande do Sul}
\resizebox{1.0\textwidth}{!}{%
\begin{tabular}{lccc} \toprule
$\Delta$ between 1950 and 1940 Census (N of) & Literate & Completed Primary School & Completed High School \\ \hline
Non-Southern European Immigrants & 7.086*** & 7.678*** & 1.807*** \\
Schools in rural areas & 379.189*** & 518.163*** & 136.898*** \\
\textbf{Inter. Rural schools * Non-South. Eur. Imm.} & \textbf{-0.314***} & \textbf{-0.438***} & \textbf{-0.119***} \\
 &  &  &   \\
N of observations & 86 & 86 & 86 \\
Controls at the municipal level & X & X & X \\
Mean Dependent Variable & 4,837 & 5,066 & 547  \\
\bottomrule
\end{tabular}
}
\caption*{\footnotesize Source: Author with data from the Brazilian 1940 and 1950 Census. Ordinary Least Square results at the municipal level. Controls at the municipal level include share of economic activity by sector and share of black and brown population. There were 88 municipalities in the state of Rio Grande do Sul at the time of the 1940 Census. However, two had missing information on rural schools. *** p$<$0.01, ** p$<$0.05, * p$<$0.1. }
\label{tab:munic_results}
\end{table}

\subsection{Immigrant Participation in the Labor Market}\label{subsec:short_individual_level}

Finally, I exploit individual-level data from work card registration records in the state of Rio Grande do Sul between 1930 and 1944.\footnote{See \autoref{fig:worker_card_rs} in \autoref{sec:appendix} for an example.} These records, required for any formal employment, contain information on age, nationality, and occupation. The sample includes 1,207 registration sheets for workers of German (857) and Italian (350) origin, with approximately 75\% of entries from the state capital, Porto Alegre.

Table \ref{tab:results_drt_rs} summarizes changes in the characteristics of immigrant workers' registrations before and after the onset of the Nationalization Campaign (1938-1945). Prior to 1938, only 3\% of entries listed the worker as unemployed. During the campaign, this figure rose to 18\%, with a particularly sharp increase among Italian workers, for whom 26\% of entries indicated unemployment. The share of foreign teachers also rose by 6 percentage points, likely reflecting the displacement of immigrant educators following the closure of immigrant-community schools. Additionally, the total number of work card requests by foreign nationals declined substantially after 1938, suggesting broader disruptions in labor market integration during the period.

\bigskip

\begin{table}[H]
\centering
\caption{Requests for work card registration by immigrants in the State of Rio Grande do Sul between 1930 and 1944}
\resizebox{0.75\textwidth}{!}{%
\begin{tabular}{lcccccc} \toprule
Characteristics & \multicolumn{2}{c}{Unemployed} & \multicolumn{2}{c}{Teachers} & \multicolumn{2}{c}{Total requests (N)} \\
Time & Pre-1938 & Post-1938 & Pre-1938 & Post-1938 & Pre-1938 & Post-1938 \\ \hline
Germans & 3\% & 15\% & 0\% & 7\% & 467 & 390 \\
Italians & 3\% & 26\% & 1\% & 3\% & 209 & 141 \\
Total & 3\% & 18\% & 0\% & 6\% & 676 & 531 \\
\bottomrule
\end{tabular}
}
\caption*{\footnotesize Source: Author with data from the DRT/RS (\textit{Acervo da Delegacia Regional do Trabalho do Rio Grande do Sul}) made available by the \textit{Núcleo de Documentação Histórica – Professora Beatriz Loner, Universidade Federal de Pelotas}.}
\label{tab:results_drt_rs}
\end{table}

\section{Long-term and intergenerational effects} \label{sec:Intergenerational effects}

\noindent Consistent with evidence from other contexts \citep{fouka_2020, carvalho_koyama_2016, lleras_AEJEC_2015}, the Brazilian Nationalization Campaign had significant adverse effects on school-age immigrant children and their families. Given the strong link between intergenerational and cross-family inequality \citep{becker_tomes_1986}, such disruptions are likely to generate lasting consequences. To assess these long-term effects, I use individual-level data from the 1991, 2000, and 2010 Brazilian Censuses. Beginning in 1991, the Census recorded both immigrants’ countries of origin and their year of arrival in Brazil—key variables for the empirical strategy. These household datasets also allow the identification of second-generation immigrants, defined as individuals living with at least one foreign-born parent, enabling an intergenerational analysis of educational outcomes. The results reveal persistent gaps in attainment between immigrants, and their descendants, exposed to the Nationalization Campaign (1938–1945) and comparable cohorts of immigrant origin who completed their schooling in their countries of origin.

\subsection{Empirical strategy: Immigrant Cohorts}

At the time of the Nationalization Campaign, immigrant community schools primarily provided primary education, generally serving children aged 7 to 12. To identify individuals who were of school age during this period, I construct age-based cohorts using data from the 1991, 2000, and 2010 Brazilian Censuses. The treatment group is defined as immigrants who were between 7 and 12 years old at any point between 1938 and 1945. For instance, in the 1991 Census, this group includes individuals aged 53 (5 years old in 1945) to 65 (12 years old in 1938). See Table \ref{tab:census_ages_treat} in Appendix \ref{sec:appendix} for the complete age brackets. Consistent with existing literature on skill formation \citep{heckman_Science_2006}, we would expect the effects of the campaign to be stronger among those who were younger at the time.

This strategy rests on the assumption that immigrants who arrived in Brazil before 1945, thus potentially exposed to the Nationalization Campaign, are comparable to those who arrived shortly thereafter and were not subject to the policy during their formative schooling years. By holding constant both age at arrival and country of origin, I compare cohorts that differ only in their year of arrival, before or after the campaign, to estimate its long-term effects of the assimilation policies. Immigrants arriving after 1945 serve as a plausible counterfactual group, representing the likely outcomes for those who were exposed, had they not been subjected to the NC. I examine heterogeneous effects by country of origin, using cultural proximity as a guiding framework.

To define treatment and control groups, I use shifts in immigration legislation as cutoffs: the treated group includes immigrants who arrived between 1930 and 1945, while the comparison group includes those who arrived between 1946 and 1963.\footnote{See Section \ref{sec:Historical Context} for further details.} The sample includes school-age immigrants and their household members observed in the 1991, 2000, and 2010 Censuses. I estimate reduced-form effects of the Nationalization Campaign on both directly affected individuals and their descendants using the following specification:

    \begin{equation*}
    \begin{aligned}
        {y}_{i,t} = \alpha + \beta \; \mathbbm{1}_{1930-1945} + {X'}_{i,t} \; \Gamma + \delta_{m} + \epsilon_{i,t}
    \end{aligned}
    \end{equation*}

where \textit{y} denotes the long-term outcome of individual \textit{i}, either an immigrant or their descendant, observed in Census year \textit{t} (1991, 2000, or 2010). The indicator variable $\mathbbm{1}_{1930-1945}$ equals one for individuals who arrived in Brazil before 1945, and for household members related to them, thereby identifying those exposed to the Nationalization Campaign. \textit{X'} is a matrix of individual- and household-level controls measured at the time of the Census. The term $\delta_{m}$ captures municipality fixed effects, and $\epsilon_{i,t}$ is the robust, idiosyncratic error term.

This analysis is subject to some limitations. First, it includes only individuals who survived until the Census years, which may introduce survivorship bias. Second, it does not capture Brazilian-born children of immigrants who may have attended community schools at the time of the NC but are not identifiable as part of the immigrant population in the data, though they are not part of the comparison groups in this analysis. Finally, the intergenerational analysis is restricted to cases in which the adult children of immigrants continued to reside in the same household at the time of the Census.


    


\subsection{Baseline characteristics}

Before presenting the reduced-form effects of the Nationalization Campaign on the long-term and intergenerational outcomes of immigrants and their descendants, I first assess whether families with school-age children differed in observable characteristics before and after 1945. To do so, I randomly selected approximately 100 passenger lists from ships arriving at the port of Santos between 1930 and 1963, ensuring a balanced distribution across years. From these lists, I extracted information on each permanent immigrant, organized by family units.

Table \ref{tab:ttest_imm_arrival} presents the characteristics of fathers and mothers of school-age children (7–12 years old) from non-Southern European countries. There are no significant differences in literacy rates between parents who arrived before and after 1945. Although marginally significant, a smaller share of fathers arriving after 1945 were employed as farmers, possibly reflecting Brazil's industrial development during the postwar years.\footnote{Historical evidence also suggests that many farmers transitioned to urban employment after 1945.} Since parental literacy and occupation, especially that of fathers, are strong predictors of children's educational attainment \citep{abramitzky_2021}, these similarities support the comparability of the two groups. Finally, a greater share of post-1945 arrivals identified as Catholic, though this difference was only marginally significant among mothers.

\bigskip

\begin{table}[H]
\centering
\caption{Comparison of characteristics of immigrants from non-Southern European countries in families with children at school age (7-12 years old) at arrival in the Port of Santos}
\resizebox{0.55\textwidth}{!}{%
\begin{tabular}{lccc} \toprule
Average Characteristic & Before 1945 & After 1945 & Diff. \\ \hline
Father was literate & 1.00 & 0.99 & 0.01 \\
Father was a farmer & 0.91 & 0.83 & 0.08* \\
Father was catholic & 0.06 & 0.10 & -0.04 \\
Mother was literate & 1.00 & 0.98 & 0.02 \\
Mother was catholic & 0.04 & 0.11 & -0.07* \\
 &  &  &  \\
N of fathers & 107 & 117 & - \\
N of mothers & 104 & 112 & - \\ \bottomrule
\end{tabular}
}
\caption*{\footnotesize Source: Author with data from the passengers lists of immigrants arriving at the port of Santos between 1930 and 1963. *** p$<$0.01, ** p$<$0.05, * p$<$0.1. }
\label{tab:ttest_imm_arrival}
\end{table}

Another important consideration is the distribution of immigrants by origin within the non-Southern European group, across both data sources and time of arrival. Table \ref{tab:immig_origin_census} shows the top four origins of school-age non-Southern European immigrants during the Nationalization Campaign, based on the Brazilian Census and passenger lists from the Port of Santos.\footnote{The Port of Santos is one of Brazil's main entry points, though not the only one, so some discrepancies between the two sources are to be expected.} The distribution is relatively consistent, though immigrants of Japanese origin made up a substantially larger share before 1945. To address this, I also demonstrate that the main results hold when analyzing each country of origin separately. While some estimates may be less precise due to smaller sample sizes, the overall group-level patterns remain robust.

\bigskip

\begin{table}[H]
\centering
\caption{Top four origins of non-Southern European immigrants in school-age at the time of the Nationalization Campaign}
\resizebox{0.95\textwidth}{!}{%
\begin{tabular}{lcccccccc} \toprule
Country of Origin & \multicolumn{2}{c}{Germany} & \multicolumn{2}{c}{Japan} & \multicolumn{2}{c}{Lebanon} & \multicolumn{2}{c}{Poland} \\ \hline
Arrival & before 1945 & after 1945 & b. 1945 & a. 1945 & b. 1945 & a. 1945 & b. 1945 & a. 1945 \\ \hline
1991 Census & 7\% & 9\% & 64\% & 34\% & 4\% & 13\% & 8\% & 4\% \\
2000 Census & 6\% & 9\% & 71\% & 37\% & 2\% & 12\% & 8\% & 4\% \\
2010 Census & 5\% & 9\% & 73\% & 41\% & 2\% & 13\% & 8\% & 2\% \\
Passengers lists & 10\% & 10\% & 76\% & 71\% & 1\% & 4\% & 1\% & 1\% \\ \bottomrule
\end{tabular}
}
\caption*{\footnotesize Source: Author with data from the Brazilian 1991, 2000, and 2010 Censuses; and passengers lists of immigrants arriving at the port of Santos between 1930 and 1963. The group of non-Southern European immigrants are those from Asian and European countries, except Italy, Spain, or Portugal.}
\label{tab:immig_origin_census}
\end{table}

\subsection{Results: Long-term and intergenerational}

Table \ref{tab:results_long_ind} presents results of OLS estimations with individual- and household-level controls, and municipality fixed effects for individual immigrants in school age who arrived before and after 1945 in the Brazilian Censuses. The estimates show a negative association for non-Southern European immigrants who were of school age during the Nationalization Campaign, relative to those who arrived afterward. While it is not possible to track individuals across the 1991, 2000, and 2010 Censuses, it is likely that many are observed in multiple waves, given the age-based sample.\footnote{The decline in sample size over time reflects natural attrition, as the oldest individuals would have reached age 84 by 2010.}

The significance of these results lies in the fact that, decades after the Nationalization Campaign, immigrants who were subjected to assimilation policies during their primary school years were 12\% less likely to complete high school, 6\% less likely to attain a college education, and 4\% less likely to complete primary education as the 1991 Census. On average, they had 1.7 fewer years of education. Surprisingly, these immigrants were more likely to be literate, 3\% and to identify as Catholic, 4\%. These effects are especially pronounced among Japanese immigrants, as shown in Table \ref{tab:results_long_ind_japanese}.

These findings suggest long-lasting negative effects of the forced assimilation policy on school-age children, which resulted in poorer human capital formation, as evidenced by lower educational attainment across their lifetime. While the mechanisms remain unclear, we observe that those who were exposed to Brazilian culture earlier in life, particularly through the Nationalization Campaign, were more likely to convert to Catholicism. The mandatory integration into the Brazilian public education system may have played a key role in this outcome.

\bigskip

\begin{table}[H]
\centering
\caption{Long-term results - Individual immigrants from non-Southern European origins who arrived before and after 1945 in school age during the Nationalization Campaign}
\resizebox{0.98\textwidth}{!}{%
\begin{tabular}{lcccccc} \toprule
Ind. characteristics & Primary Education & High School & College Education & Years of Education & Literate & Catholic \\ \hline
 & \multicolumn{6}{c}{\textit{Immigrants from non-Southern European origins in the 1991 Census}} \\
Arrival $\mathbbm{1}_{1930-1945}$ & \textbf{-0.038***} & \textbf{-0.116***} & \textbf{-0.059***} & \textbf{-1.715***} & \textbf{0.033***} & \textbf{0.042**} \\
N of observations & 4,665 & 4,665 & 4,665 & 4,665 & 4,665 & 4,665 \\
 &  &  &  &  &  &  \\
 & \multicolumn{6}{c}{\textit{Immigrants from non-Southern European origins in the 2000 Census}} \\
Arrival $\mathbbm{1}_{1930-1945}$ & \textbf{-0.037**} & \textbf{-0.147***} & \textbf{-0.048***} & \textbf{-1.691***} & \textbf{0.040***} & \textbf{0.108***} \\
N of observations & 4,289 & 4,289 & 4,289 & 4,289 & 4,289 & 4,289 \\
 &  &  &  &  &  &  \\
 & \multicolumn{6}{c}{\textit{Immigrants from non-Southern European origins in the 2010 Census}} \\
Arrival $\mathbbm{1}_{1930-1945}$ & -0.025 & \textbf{-0.092***} & \textbf{-0.073**} & - & \textbf{0.062***} & \textbf{0.083**} \\
N of observations & 2,010 & 2,010 & 2,010 & - & 2,010 & 2,010 \\ \bottomrule
\end{tabular}
}
\caption*{\footnotesize Source: Author, by using microdata from individuals and households in the 1991, 2000, and 2010 Brazilian Censuses (Datacem project). Estimates are based on OLS regressions with individual- and household-level controls and municipality fixed effects. Regressions use Census sample weights and include migrants of non-Southern European origin who arrived before or after 1945 and were of school age (7–12 years old) during the Nationalization Campaign. *** p$<$0.01, ** p$<$0.05, * p$<$0.1.}
\label{tab:results_long_ind}
\end{table}

I now turn to the intergenerational dimension of human capital formation. Table \ref{tab:results_long_son} presents results for the second generation, sons and daughters of non-Southern European immigrants who arrived before or after 1945, focusing on the same age cohorts as in the previous analysis. As expected, the lower educational attainment of first-generation immigrants exposed to the Nationalization Campaign is reflected in their children's outcomes. Despite substantial improvements in Brazil's education system over time, children of exposed parents were, on average, 3\% less likely to complete college and had 0.6 fewer years of schooling by the 1991 Census. This analysis is restricted to children who cohabited with their immigrant parents at the time of the survey. In the 2000 Census, although sample selection becomes more pronounced, the estimated effects are even larger: college completion was 7\% lower, and average schooling was reduced by 1.2 years.

Another notable finding is the increased likelihood of religious conversion among children of earlier-arriving immigrants. Compared to those who arrived after 1945, these children were significantly more likely to identify as Catholic: by 6\%, 13\%, and 18\% in the 1991, 2000, and 2010 Censuses, respectively. While the precise mechanisms remain uncertain, it is plausible that exposure to the Nationalization Campaign contributed to this shift, particularly given Catholicism's status as Brazil's dominant religion and its role as a political ally of Vargas's regime. The effect is especially pronounced among Japanese immigrants, as shown in Table \ref{tab:results_long_son_japanese}. The takeaway of these analyses is that policy decisions made today can have lasting effects on individuals for generations to come.

\begin{table}[H]
\centering
\caption{Long-term results - Sons and daughters of immigrants from non-Southern European origins who arrived before and after 1945 in school age during the Nationalization Campaign}
\resizebox{1.0\textwidth}{!}{%
\begin{tabular}{lcccccc} \toprule
Ind. characteristics & Primary Education & High School & College Education & Years of Education & Literate & Catholic \\ \hline
 & \multicolumn{6}{c}{\textit{Sons and daughters of immigrants from non-Southern European origins in the 1991 Census}} \\
Arrival $\mathbbm{1}_{1930-1945}$ & 0.001 & 0.016 & \textbf{-0.033*} & \textbf{-0.567***} & \textbf{-0.009} & \textbf{0.058***} \\
N of observations & 4,869 & 4,869 & 4,869 & 4,869 & 4,869 & 4,869 \\
 &  &  &  &  &  &  \\
 & \multicolumn{6}{c}{\textit{Sons and daughters of immigrants from non-Southern European origins in the 2000 Census}} \\
Arrival $\mathbbm{1}_{1930-1945}$ & -0.012 & 0.020 & \textbf{-0.070***} & \textbf{-1.238***} & \textbf{-0.040***} & \textbf{0.133***} \\
N of observations & 2,799 & 2,799 & 2,799 & 2,799 & 2,799 & 2,799 \\
 &  &  &  &  &  &  \\
 & \multicolumn{6}{c}{\textit{Sons and daughters of immigrants from non-Southern European origins in the 2010 Census}} \\
Arrival $\mathbbm{1}_{1930-1945}$ & 0.073* & 0.002 & \textbf{-0.081} & - & \textbf{0.002} & \textbf{0.184***} \\
N of observations & 828 & 828 & 828 & - & 828 & 828 \\ \bottomrule
\end{tabular}
}
\caption*{\footnotesize Source: Author, by using microdata from individuals and households in the 1991, 2000, and 2010 Brazilian Censuses (Datacem project). Estimates are based on OLS regressions with individual- and household-level controls and municipality fixed effects. Regressions use Census sample weights and include sons and daughters of immigrants from non-Southern European origins who arrived before or after 1945 and were of school age (7–12 years old) during the Nationalization Campaign, who cohabit with them. *** p$<$0.01, ** p$<$0.05, * p$<$0.1.}
\label{tab:results_long_son}
\end{table}

\subsection{Heterogeneous Effects by Educational Level}

Although the primary focus has been on the closure of immigrant-community schools, primarily primary education institutions, it is important to acknowledge that I cannot confirm whether the immigrants observed in the 1991, 2000, and 2010 Censuses actually attended these schools. There may have been variation both in the educational levels offered by those schools and in the age at which children attended them, given the age brackets used here. Nonetheless, the literature on skill formation suggests that the impact of educational disruptions varies across stages of childhood \citep{cunha_heckman_AER_2007}. In this context, we would expect heterogeneous effects of the Nationalization Campaign depending on a child's age and educational stage at the time. \citet{heckman_Science_2006} emphasizes that policy interventions can have asymmetric effects depending on when they occur in the life cycle, particularly in shaping skills.

To explore this heterogeneity, I conduct a robustness check by dividing individuals into groups corresponding to primary, secondary, and tertiary school ages during the campaign. The underlying hypothesis is that older children may also have been affected, though for shorter or different durations. Children aged 12–15 in 1945 are assumed to have been in secondary school, and those aged 15–18 in 1945 in tertiary education. For the full age brackets, see Table \ref{tab:census_ages_treat} in Appendix \ref{sec:appendix}.\footnote{Age groups may overlap due to age-grade distortion and regional differences in education laws.}

Table \ref{tab:results_long_educ_levels} presents results by school level, based on individuals' age during the campaign. The negative effects on educational attainment are either decreasing with age or remain constant across educational stages. Notably, among those who completed high school, the primary educational goal for most of this generation, the largest effects are observed for individuals who were of primary-school age during the campaign, followed by smaller but still significant effects for those in the secondary and tertiary age groups. These findings align with the literature linking early educational opportunities to long-term human capital accumulation and how disruptive the assimilation policies were for the younger cohort.

\begin{table}[H]
\centering
\caption{Long-term results - Individual migrants from non-Southern European origins who arrived before and after 1945 by school levels during the Nationalization Campaign}
\resizebox{1.0\textwidth}{!}{%
\begin{tabular}{lccccccc} \toprule
Individual characteristics & Primary Education & High School & College Education & Years of Education & Literate & Catholic & N \\ \hline
Arrival $\mathbbm{1}_{1930-1945}$ & \multicolumn{7}{c}{\textit{Immigrants from non-Southern European origins in the 1991 Census}} \\
 $  $ Primary Education (7-12 yo) & \textbf{-0.038***} & \textbf{-0.116***} & -0.059*** & -1.715*** & 0.033*** & 0.042** & 4,665 \\
 $  $ Secondary Education (12-15 yo) & \textbf{-0.027**} & \textbf{-0.095***} & -0.061*** & -1.542*** & \multicolumn{1}{l}{0.039***} & 0.005 & 4,590 \\
 $  $ Tertiary Education (15-18 yo) & \textbf{-0.017} & \textbf{-0.083***} & -0.059*** & -1.478*** & 0.013 & -0.006 & 4,559 \\
  &  &  &  &  &  & & \\
Arrival $\mathbbm{1}_{1930-1945}$ & \multicolumn{7}{c}{\textit{Immigrants from non-Southern European origins in the 2000 Census}} \\
 $  $ Primary Education (7-12 yo) & \textbf{-0.037**} & \textbf{-0.147***} & -0.048*** & -1.691*** & 0.040*** & 0.108*** & 4,289 \\
 $  $ Secondary Education (12-15 yo) & \textbf{-0.038**} & \textbf{-0.116***} & -0.083*** & -1.976*** & 0.042*** & 0.056*** & 3,999 \\
 $  $ Tertiary Education (15-18 yo) & \textbf{-0.031*} & \textbf{-0.101***} & -0.088*** & -1.951*** & 0.033** & 0.010 & 3,726 \\
  &  &  &  &  &  & & \\
Arrival $\mathbbm{1}_{1930-1945}$ & \multicolumn{7}{c}{\textit{Immigrants from non-Southern European origins in the 2010 Census}} \\
 $  $ Primary Education (7-12 yo) & \textbf{-0.025} & \textbf{-0.092***} & -0.073** & - & 0.062*** & 0.083** & 2,010 \\
 $  $ Secondary Education (12-15 yo) & \textbf{-0.017} & \textbf{-0.059*} & -0.092*** & - & 0.058*** & 0.070* & 1,658 \\
 $  $ Tertiary Education (15-18 yo) & \textbf{-0.006} & \textbf{-0.040} & -0.096*** & - & 0.048* & -0.011 & 1,375 \\ \bottomrule
\end{tabular}
}
\caption*{\footnotesize Source: Author, by using microdata from individuals and households in the 1991, 2000, and 2010 Brazilian Censuses (Datacem project). Estimates are based on OLS regressions with individual- and household-level controls and municipality fixed effects. Regressions use Census sample weights and include migrants of non-Southern European origin who arrived before or after 1945 and were of school age (7–12 years old) during the Nationalization Campaign. *** p$<$0.01, ** p$<$0.05, * p$<$0.1.}
\label{tab:results_long_educ_levels}
\end{table}

\subsection{Mechanism: Cultural Proximity}

I now turn to the main mechanism proposed to explain the effects of the assimilation policy on educational outcomes: cultural proximity between immigrants' countries of origin and Brazil. Given the official rhetoric during the period, immigrants from Axis countries (Italy, Germany, and Japan) were among those most directly targeted by institutional measures (see \autoref{fig:newspapers_nc} in \autoref{sec:appendix}). To explore this, I compare the long-term, individual-level outcomes for immigrants from each of these countries. The results indicate that, holding the intensity of assimilation policies constant, immigrants from more culturally distant origins experienced the largest long-term effects.

\subsubsection{Italian immigrants}

The case of Italian immigrants also serves as a robustness check for the main long-term results. If the effects were driven primarily by the poor quality of Brazil’s educational system, we would expect similar patterns among immigrants from countries with comparable systems, such as Italy. In fact, the educational system in many Italian regions, especially those with high emigration rates, was similarly developed to Brazil's and less advanced than that of other Axis countries. Although Italians were also affected by other aspects of the Nationalization Campaign, including labor market restrictions (see Subsection \ref{subsec:short_individual_level}), given a cultural proximity between Brazil and Southern Europe, we would not expect significant long-term effects on educational or cultural outcomes for Italian immigrants.

Table \ref{tab:results_long_ind_italian} presents the results for individual Italian immigrants. Overall, there are few significant differences in educational and religious outcomes between those who arrived before or after the Nationalization Campaign. Most effects are either positive or close to zero. When statistically significant, the results indicate that pre-1945 arrivals were more likely to enter the education system, complete college, and attain more years of schooling—contrasting with earlier findings for the non-Southern European immigrant groups. These patterns suggest that, although Italians were also targeted by the assimilation policies, their high cultural proximity to Brazil mitigated its negative effects on educational and religious outcomes.

\begin{table}[H]
\centering
\caption{Long-term results - Individual Italian immigrants who arrived before and after 1945 in school age during the Nationalization Campaign}
\resizebox{1.00\textwidth}{!}{%
\begin{tabular}{lcccccc} \toprule
Individual characteristics & Primary Education & High School & College Education & Years of Education & Literate & Catholic \\ \hline
 & \multicolumn{6}{c}{\textit{Immigrants from Italy in the 1991 Census}} \\
Arrival $\mathbbm{1}_{1930-1945}$ & -0.003 & -0.011 & 0.068*** & 0.754*** & 0.006 & -0.008 \\
N of observations & 1,862 & 1,862 & 1,862 & 1,862 & 1,862 & 1,862 \\
 &  &  &  &  &  &  \\
 & \multicolumn{6}{c}{\textit{Immigrants from Italy in the 2000 Census}} \\
Arrival $\mathbbm{1}_{1930-1945}$ & 0.002 & 0.013 & 0.051 & 0.848** & 0.010 & -0.023 \\
N of observations & 1,733 & 1,733 & 1,733 & 1,733 & 1,733 & 1,733 \\
 &  &  &  &  &  &  \\
 & \multicolumn{6}{c}{\textit{Immigrants from Italy in the 2010 Census}} \\
Arrival $\mathbbm{1}_{1930-1945}$ & -0.035 & 0.099 & 0.106 & - & 0.076** & -0.045 \\
N of observations & 692 & 692 & 692 & - & 692 & 692 \\ \bottomrule
\end{tabular}
}
\caption*{\footnotesize Source: Author, by using microdata from individuals and households in the 1991, 2000, and 2010 Brazilian Censuses (Datacem project). Estimates are based on OLS regressions with individual- and household-level controls and municipality fixed effects. Regressions use Census sample weights and include immigrants from Italy who arrived before or after 1945 and were of school age (7–12 years old) during the Nationalization Campaign. *** p$<$0.01, ** p$<$0.05, * p$<$0.1.}
\label{tab:results_long_ind_italian}
\end{table}

The second generation analysis, sons and daughters of Italian immigrants, shows a slightly different pattern. Table \ref{tab:results_long_son_italian} presents results for those cohabiting with their parents at the time of the Censuses. While labor market restrictions and cultural suppression during the Nationalization Campaign may have affected their college completion and years of education through socioeconomic hardship, most other outcomes are statistically insignificant or close to zero.

\begin{table}[H]
\centering
\caption{Long-term results - Sons and daughters of Italy immigrants who arrived before and after 1945 in school age during the Nationalization Campaign}
\resizebox{1.0\textwidth}{!}{%
\begin{tabular}{lcccccc} \toprule
Individual characteristics & Primary Education & High School & College Education & Years of Education & Literate & Catholic \\ \hline
 & \multicolumn{6}{c}{\textit{Sons and daughters of immigrants from Italy in the 1991 Census}} \\
Arrival $\mathbbm{1}_{1930-1945}$ & 0.033* & 0.001 & -0.069*** & -0.721*** & -0.009 & 0.034* \\
N of observations & 1,939 & 1,939 & 1,939 & 1,939 & 1,939 & 1,939 \\
 &  &  &  &  &  &  \\
 & \multicolumn{6}{c}{\textit{Sons and daughters of immigrants from Italy in the 2000 Census}} \\
Arrival $\mathbbm{1}_{1930-1945}$ & 0.025 & 0.025 & -0.046 & -0.691 & -0.013 & 0.033 \\
N of observations & 1,123 & 1,123 & 1,123 & 1,123 & 1,123 & 1,123 \\
 &  &  &  &  &  &  \\
 & \multicolumn{6}{c}{\textit{Sons and daughters of immigrants from Italy in the 2010 Census}} \\
Arrival $\mathbbm{1}_{1930-1945}$ & 0.029 & -0.055 & -0.032 & - & 0.034 & 0.141 \\
N of observations & 255 & 255 & 255 & - & 255 & 255 \\ \bottomrule
\end{tabular}
}
\caption*{\footnotesize Source: Author, by using microdata from individuals and households in the 1991, 2000, and 2010 Brazilian Censuses (Datacem project). Estimates are based on OLS regressions with individual- and household-level controls and municipality fixed effects. Regressions use Census sample weights and include sons and daughters of immigrants from Italy who arrived before or after 1945 and were of school age (7–12 years old) during the Nationalization Campaign, who cohabit with them. *** p$<$0.01, ** p$<$0.05, * p$<$0.1.}
\label{tab:results_long_son_italian}
\end{table}

\subsubsection{Japanese immigrants}

Japanese immigrants made up the majority of non-Southern European arrivals between 1930 and 1963 (\autoref{tab:immig_origin_census}) and were also the most culturally distant major immigrant group in Brazil (\autoref{tab:language_home_1940}). As such, much of the variation observed in earlier results likely reflects the particularly disruptive effects of the assimilation policies on Japanese communities.

\autoref{tab:results_long_ind_japanese} shows large and significant long-term effects of these policies on the educational and religious outcomes of individual Japanese immigrants. In the 1991 Census, those exposed to the Nationalization Campaign were, on average, 6\% less likely to complete primary school, 16\% less likely to finish high school, 8\% less likely to complete college, had 2.6 fewer years of education, and were 11\% more likely to identify as Catholic in adulthood. While some of the religious change may reflect baseline group differences (\autoref{tab:ttest_imm_arrival}), the magnitude of the educational effects underscores the severe impact of the campaign on this group.

\begin{table}[H]
\centering
\caption{Long-term results - Individual Japanese immigrants who arrived before and after 1945 in school age during the Nationalization Campaign}
\resizebox{1.00\textwidth}{!}{%
\begin{tabular}{lcccccc}
\hline
Individual characteristics & Primary Education & High School & College Education & Years of Education & Literate & Catholic \\ \hline
 & \multicolumn{6}{c}{\textit{Immigrants from Japan in the 1991 Census}} \\
Arrival $\mathbbm{1}_{1930-1945}$ & -0.058*** & -0.161*** & -0.084*** & -2.594*** & 0.039** & 0.111*** \\
N of observations & 2,293 & 2,293 & 2,293 & 2,293 & 2,293 & 2,293 \\
 &  &  &  &  &  &  \\
 & \multicolumn{6}{c}{\textit{Immigrants from Japan in the 2000 Census}} \\
Arrival $\mathbbm{1}_{1930-1945}$ & -0.051** & -0.175*** & -0.076*** & -2.378*** & 0.068*** & 0.172*** \\
N of observations & 2,258 & 2,258 & 2,258 & 2,258 & 2,258 & 2,258 \\
 &  &  &  &  &  &  \\
 & \multicolumn{6}{c}{\textit{Immigrants from Japan in the 2010 Census}} \\
Arrival $\mathbbm{1}_{1930-1945}$ & -0.026 & -0.164*** & -0.080** & - & 0.104*** & 0.175*** \\
N of observations & 1,172 & 1,172 & 1,172 & - & 1,172 & 1,172 \\ \bottomrule
\end{tabular}
}
\caption*{\footnotesize Source: Author, by using microdata from individuals and households in the 1991, 2000, and 2010 Brazilian Censuses (Datacem project). Estimates are based on OLS regressions with individual- and household-level controls and municipality fixed effects. Regressions use Census sample weights and include migrants from Japan who arrived before or after 1945 and were of school age (7–12 years old) during the Nationalization Campaign. *** p$<$0.01, ** p$<$0.05, * p$<$0.1.}
\label{tab:results_long_ind_japanese}
\end{table}

Similarly, \autoref{tab:results_long_son_japanese} presents the results for the second generation of Japanese immigrants in Brazil, which closely mirror the patterns observed for the first generation. While estimates vary across Censuses, the 2000 Census shows that sons and daughters of Japanese immigrants exposed to the campaign were, on average, 5\% less likely to complete primary school, 17\% less likely to finish high school, 8\% less likely to complete college, had 2.4 fewer years of education, and were 17\% more likely to identify as Catholic in adulthood. These results suggest that cultural distance significantly amplified the long-term impacts of the assimilation policies.

\begin{table}[H]
\centering
\caption{Long-term results - Sons and daughters of Japanese immigrants who arrived before and after 1945 in school age during the Nationalization Campaign}
\resizebox{1.00\textwidth}{!}{%
\begin{tabular}{lcccccc} \toprule
Individual characteristics & Primary Education & High School & College Education & Years of Education & Literate & Catholic \\ \hline
 & \multicolumn{6}{c}{\textit{Sons and daughters of immigrants from Japan in the 1991 Census}} \\
Arrival $\mathbbm{1}_{1930-1945}$ & 0.009 & -0.008 & 0.014 & -0.467** & -0.013 & 0.103*** \\
N of observations & 2,609 & 2,609 & 2,609 & 2,609 & 2,609 & 2,609 \\
 &  &  &  &  &  &  \\
 & \multicolumn{6}{c}{\textit{Sons and daughters of immigrants from Japan in the 2000 Census}} \\
Arrival $\mathbbm{1}_{1930-1945}$ & -0.051** & -0.175*** & -0.076*** & -2.378*** & 0.068*** & 0.172*** \\
N of observations & 2,258 & 2,258 & 2,258 & 2,258 & 2,258 & 2,258 \\
 &  &  &  &  &  &  \\
 & \multicolumn{6}{c}{\textit{Sons and daughters of immigrants from Japan in the 2010 Census}} \\
Arrival $\mathbbm{1}_{1930-1945}$ & 0.080 & -0.067 & 0.018 & - & 0.011 & 0.139 \\
N of observations & 553 & 553 & 553 & - & 553 & 553 \\ \bottomrule
\end{tabular}
}
\caption*{\footnotesize Source: Author, by using microdata from individuals and households in the 1991, 2000, and 2010 Brazilian Censuses (Datacem project). Estimates are based on OLS regressions with individual- and household-level controls and municipality fixed effects. Regressions use Census sample weights and include sons and daughters of immigrants from Japan who arrived before or after 1945 and were of school age (7–12 years old) during the Nationalization Campaign, who cohabit with them. *** p$<$0.01, ** p$<$0.05, * p$<$0.1.}
\label{tab:results_long_son_japanese}
\end{table}

\subsubsection{German immigrants}

In comparison, \autoref{tab:results_long_ind_german} reports results for German immigrants. Although the small sample limits estimate precision, the findings align with expectations given the cultural distance between Germany and Brazil. When significant, arriving before 1945, thus being exposed to assimilation policies, is associated with lower educational outcomes. In contrast, for religious affiliation, there appears to be a backlash effect: earlier arrivals are less likely to identify as Catholic.

\begin{table}[H]
\centering
\caption{Long-term results - Individual German immigrants who arrived before and after 1945 in school age during the Nationalization Campaign}
\resizebox{1.00\textwidth}{!}{%
\begin{tabular}{lcccccc}
\hline
Individual characteristics & Primary Education & High School & College Education & Years of Education & Literate & Catholic \\ \hline
 & \multicolumn{6}{c}{\textit{Immigrants from Germany in the 1991 Census}} \\
Arrival $\mathbbm{1}_{1930-1945}$ & 0.064 & -0.056 & -0.118 & -1.431** & 0.000 & -0.093 \\
N of observations & 372 & 372 & 372 & 372 & 372 & 372 \\
 &  &  &  &  &  &  \\
 & \multicolumn{6}{c}{\textit{Immigrants from Germany in the 2000 Census}} \\
Arrival $\mathbbm{1}_{1930-1945}$ & -0.017 & -0.182** & 0.061 & -0.470 & 0.009 & -0.065 \\
N of observations & 329 & 329 & 329 & 329 & 329 & 329 \\
 &  &  &  &  &  &  \\
 & \multicolumn{6}{c}{\textit{Immigrants from Germany in the 2010 Census}} \\
Arrival $\mathbbm{1}_{1930-1945}$ & -0.111 & 0.088 & -0.068 & - & 0.000* & -0.458*** \\
N of observations & 154 & 154 & 154 & - & 154 & 154 \\ \bottomrule
\end{tabular}
}
\caption*{\footnotesize Source: Author, by using microdata from individuals and households in the 1991, 2000, and 2010 Brazilian Censuses (Datacem project). Estimates are based on OLS regressions with individual- and household-level controls and municipality fixed effects. Regressions use Census sample weights and include migrants from Germany who arrived before or after 1945 and were of school age (7–12 years old) during the Nationalization Campaign. *** p$<$0.01, ** p$<$0.05, * p$<$0.1.}
\label{tab:results_long_ind_german}
\end{table}

Less can be concluded about the second generation of Germans in Brazil due to an even smaller sample size. While some educational outcomes show a negative association with exposure to assimilation policies, college completion appears positively correlated. Regarding religion, there is a consistent negative effect.

\begin{table}[H]
\centering
\caption{Long-term results - Sons and daughters of German immigrants who arrived before and after 1945 in school age during the Nationalization Campaign}
\resizebox{1.00\textwidth}{!}{%
\begin{tabular}{lcccccc} \toprule
Individual characteristics & Primary Education & High School & College Education & Years of Education & Literate & Catholic \\ \hline
 & \multicolumn{6}{c}{\textit{Sons and daughters of immigrants from Germany in the 1991 Census}} \\
Arrival $\mathbbm{1}_{1930-1945}$ & -0.071 & -0.085 & 0.054 & 0.587 & 0.029 & -0.102 \\
N of observations & 267 & 267 & 267 & 267 & 267 & 267 \\
 &  &  &  &  &  &  \\
 & \multicolumn{6}{c}{\textit{Sons and daughters of immigrants from Germany in the 2000 Census}} \\
Arrival $\mathbbm{1}_{1930-1945}$ & -0.002 & -0.004 & 0.082 & -0.616 & -0.010 & -0.162 \\
N of observations & 148 & 148 & 148 & 148 & 148 & 148 \\
 &  &  &  &  &  &  \\
 & \multicolumn{6}{c}{\textit{Sons and daughters of immigrants from Germany in the 2010 Census}} \\
Arrival $\mathbbm{1}_{1930-1945}$ & 1.000*** & 0.146 & -1.000*** & - & -0.000 & -1.000*** \\
N of observations & 32 & 32 & 32 & - & 32 & 32 \\ \bottomrule
\end{tabular}
}
\caption*{\footnotesize Source: Author, by using microdata from individuals and households in the 1991, 2000, and 2010 Brazilian Censuses (Datacem project). Estimates are based on OLS regressions with individual- and household-level controls and municipality fixed effects. Regressions use Census sample weights and include sons and daughters of immigrants from Germany who arrived before or after 1945 and were of school age (7–12 years old) during the Nationalization Campaign, who cohabit with them. *** p$<$0.01, ** p$<$0.05, * p$<$0.1.}
\label{tab:results_long_son_german}
\end{table}

\section{Concluding Remarks} \label{sec:Discussion and Conclusion}

\noindent This paper examines the short- and long-term effects of Brazils Nationalization Campaign (1938–1945) on the human capital development of immigrants and their descendants. The short-term evidence shows that assimilation policies negatively affected local educational outcomes, while the long-term analysis reveals persistent disadvantages for immigrants from targeted origins, particularly those from culturally distant countries such as Japan and Germany, and their children.

Drawing on multiple data sources and empirical strategies, the results are robust across specifications. Younger immigrants, especially those of primary-school age during the campaign, experienced the largest educational losses, consistent with the literature on skill formation \citep{cunha_heckman_AER_2007, heckman_Science_2006}. These effects persisted into the second generation, indicating an intergenerational transmission of disadvantage and showing how forced assimilation policies can disrupt the educational trajectories of entire families.

The findings are consistent with evidence from other historical contexts \citep{fouka_2020, carvalho_koyama_2016, lleras_AEJEC_2015} and underscore the cumulative and intergenerational impacts of early-life policy shocks. Cultural proximity played a central role: immigrants from Southern Europe, such as Italians, faced relatively smaller educational setbacks, likely due to linguistic and cultural similarities with the dominant Brazilian population. By contrast, Japanese and German immigrants, who faced greater cultural distance, suffered substantial and lasting educational penalties.

Beyond the Brazilian case, these results highlight the long-term costs of assimilation policies that dismantle community-based educational institutions. The closure of immigrant schools, combined with discrimination and exclusion from broader economic and social opportunities, hindered human capital accumulation for decades.

Finally, this study points to avenues for further research. Future work could examine the campaign’s effects on other dimensions of immigrant life—such as labor market outcomes, political participation, and social mobility—and compare them to the impacts of similar assimilation efforts in other countries and periods. Such research would deepen our understanding of how state-led assimilation shapes the long-term integration and upward mobility of immigrant communities.

\clearpage

\bibliography{references}

\clearpage

\appendix

\section{Appendix: Tables and Figures}\label{sec:appendix}

\begin{table}[H]
\centering
\caption{Effects of the Nationalization Campaign on educational outcomes given the prevalence of non-Southern European immigrants in Brazilian States}
\resizebox{1.0\textwidth}{!}{%
\begin{tabular}{lcccccc}
\hline
Dependent Variable (N) & Students & Approved Students & Schools & Private Schools & Teachers & Untrained Teachers \\ \hline
 & \multicolumn{1}{l}{} & \multicolumn{1}{l}{} & \multicolumn{1}{l}{} & \multicolumn{1}{l}{} & \multicolumn{1}{l}{} & \multicolumn{1}{l}{} \\
\multicolumn{7}{l}{\textit{Interaction of the Log number of Non-Southern European Immigrants (Continuous) with *}} \\
$ $ * $\mathbbm{1}_{1938-1945}$ (NC) & -0.009 & -0.003 & -0.012 & -0.006 & -0.017** & -0.060*** \\
$ $ * $\mathbbm{1}_{1946-1950}$ (Post-I) & -0.030** & -0.028 & -0.039** & -0.103*** & -0.038*** & -0.111*** \\
$ $ * $\mathbbm{1}_{1951-1955}$ (Post-2) & -0.035** & -0.064** & -0.065*** & -0.130*** & -0.047*** & -0.160*** \\
 & \multicolumn{1}{l}{} & \multicolumn{1}{l}{} & \multicolumn{1}{l}{} & \multicolumn{1}{l}{} & \multicolumn{1}{l}{} & \multicolumn{1}{l}{} \\
\multicolumn{7}{l}{\textit{Interaction of the binary indicator of treated states (high presence of Non-Southern European Immigrants) with *}} \\
$ $ * $\mathbbm{1}_{1938-1945}$ (NC) & -0.082** & -0.082 & -0.068 & -0.128 & -0.087 & -0.232*** \\
$ $ * $\mathbbm{1}_{1946-1950}$ (Post-I) & -0.167*** & -0.189 & -0.130 & -0.535** & -0.143 & -0.295** \\
$ $ * $\mathbbm{1}_{1951-1955}$ (Post-2) & -0.205*** & -0.379*** & -0.292*** & -0.623*** & -0.224** & -0.485*** \\
 & \multicolumn{1}{l}{} & \multicolumn{1}{l}{} & \multicolumn{1}{l}{} & \multicolumn{1}{l}{} & \multicolumn{1}{l}{} & \multicolumn{1}{l}{} \\
N of observations & 528 & 528 & 528 & 528 & 528 & 528 \\
N of unique states & 22 & 22 & 22 & 22 & 22 & 22 \\
State and Year FEs & X & X & X & X & X & X \\ \hline
\end{tabular}
}
\caption*{\footnotesize Source: Author. Results of OLS panel, difference-in-differences estimates with dynamic treatment. The treated states are those with high prevalence of non-Southern European immigrants, given by the log number of immigrants minus the number of Portuguese, Spanish, and Italian immigrants in the population of a state in the 1940 Brazilian Census. Treatment periods of time are historically defined as federal government changes. The pre-period represents the first republican rule of Getúlio Vargas; the Nationalization Campaign period, his rule as a dictator; Between 1946 and 1950, the rule of elected president Gaspar Dutra; then, Vargas again as elected president between 1951 and 1955 (although he died in 1954 and his vice took office for 1955). Educational outcome dependent variables in log number of the outcomes: private schools are used as proxies for immigrant community schools. *** p$<$0.01, ** p$<$0.05, * p$<$0.1}
\label{tab:did_short_results_ols}
\end{table}

\begin{table}[H]
\centering
\caption{Age brackets for school-age individuals between 1938-1945 observed in the 1991, 2000, and 2010 Censuses}
\resizebox{1.0\textwidth}{!}{%
\begin{tabular}{lccccc} \toprule
School level attended & Age in 1938 & Age in 1945 & Age in 1991 & Age in 2000 & Age in 2010 \\ \hline
Immigrant community schools & 7-12 & 7-12 & 53-65 & 62-74 & 72-84 \\
Primary Education & 7-12 & 7-12 & 53-65 & 62-74 & 72-84 \\
Secondary Education & 12-15 & 12-15 & 58-68 & 67-77 & 77-87 \\
Tertiary Education & 15-18 & 15-18 & 61-71 & 70-80 & 80-90 \\ \bottomrule
\end{tabular}
}
\caption*{\footnotesize Source: Author.}
\label{tab:census_ages_treat}
\end{table}

\begin{figure}[H]
    \centering
    \caption{The newspaper \textit{A Nação} (The Nation) reported the prohibition of German, Italian, and Japanese languages use in public spaces - City of Caxias do Sul, State of Rio Grande do Sul, February 1, 1942.}
    \includegraphics[width=13cm]{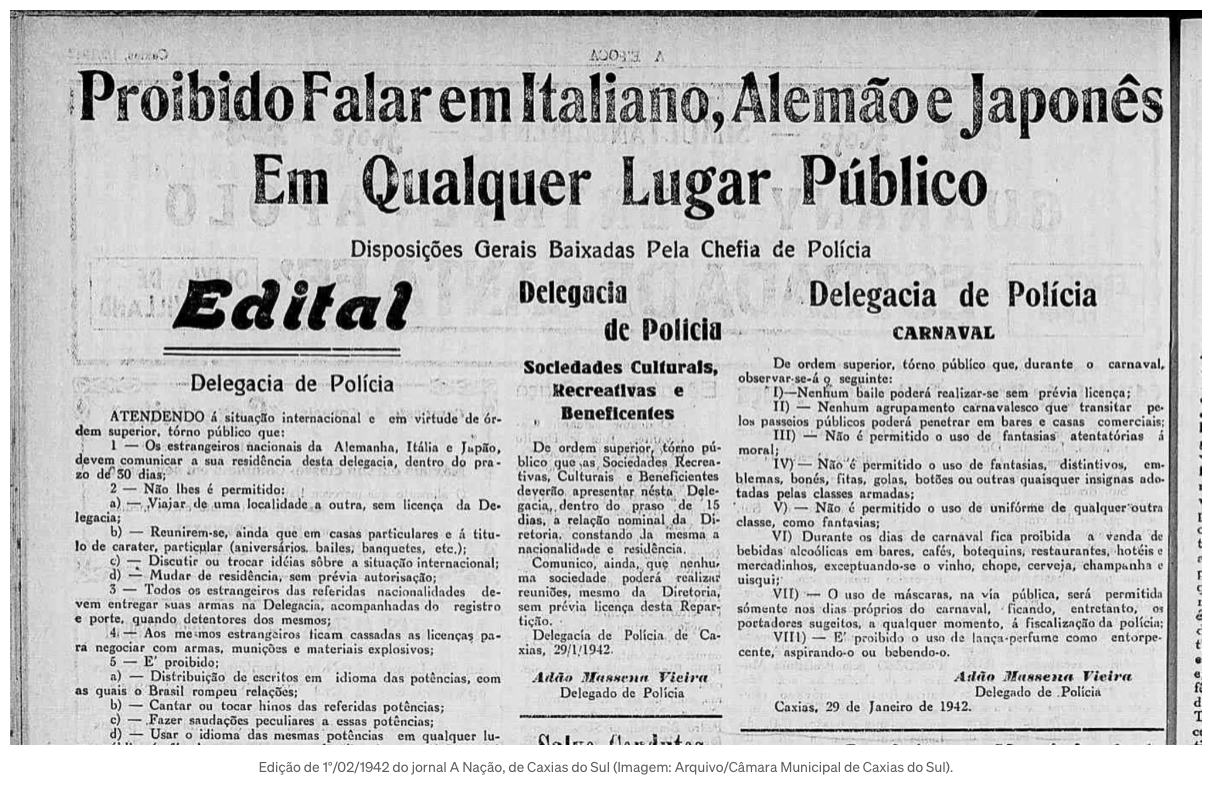}
    \label{fig:newspapers_nc}
    \caption*{\footnotesize Source: City of Caxias do Sul, Legislative archival.}
\end{figure}

\begin{figure}[H]
    \centering
    \caption{Example of a passengers list's page}
    \includegraphics[width=13cm]{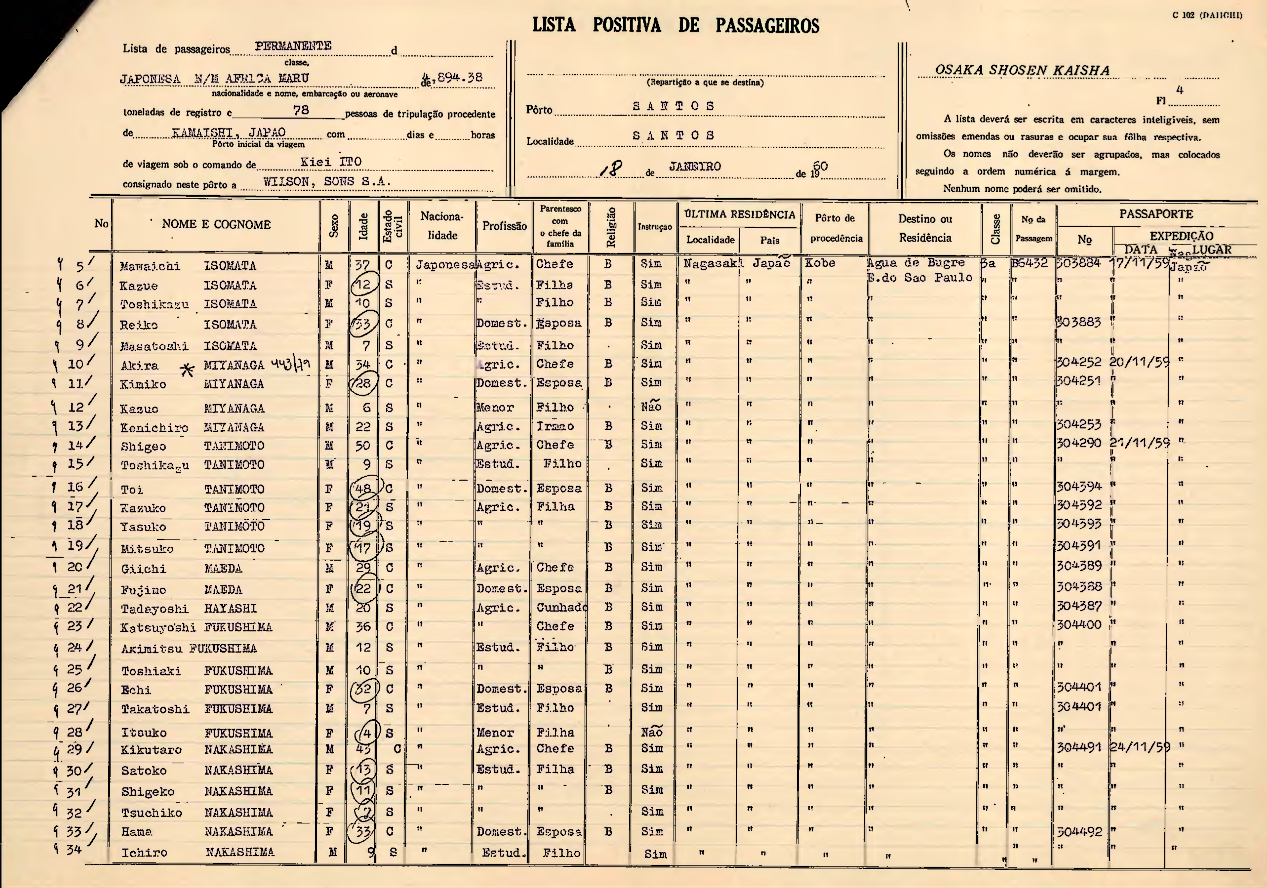}
    \label{fig:passengers_list_santos}
    \caption*{\footnotesize Source: Passengers list of ships arriving in the Port of Santos, São Paulo, Brazil.}
\end{figure}

\begin{figure}[H]
    \centering
    \caption{Example of a worker card registration sheet for the State of Rio Grande do Sul}
    \includegraphics[width=13cm]{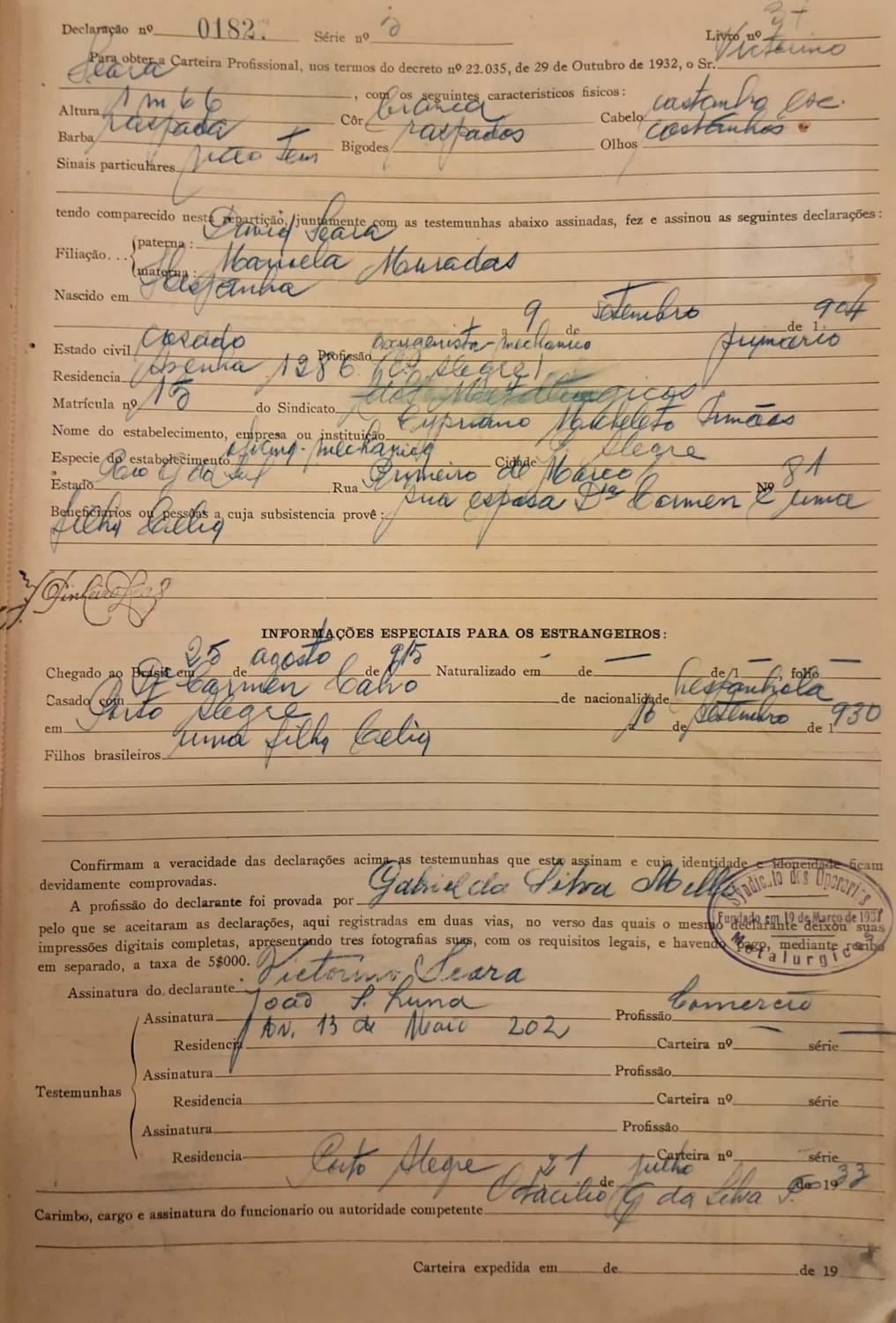}
    \label{fig:worker_card_rs}
    \caption*{\footnotesize Source: DRT/RS (\textit{Acervo da Delegacia Regional do Trabalho do Rio Grande do Sul}) made available by the \textit{Núcleo de Documentação Histórica – Professora Beatriz Loner, Universidade Federal de Pelotas}.}
\end{figure}

\end{document}